\documentclass[journal]{vgtc} 

\usepackage{cite}
\usepackage[bottom]{footmisc}
\usepackage{mathptmx}
\usepackage{graphicx}
\usepackage{siunitx}
\usepackage{times}
\usepackage[T1]{fontenc}
\usepackage{enumitem}
\usepackage[table]{xcolor} 
\usepackage{wrapfig}
\usepackage{multirow}
\usepackage{adjustbox}
\usepackage{float}
\usepackage{dirtytalk}
\usepackage{soul}
\usepackage{mdframed}
\usepackage{mwe}
\usepackage{graphbox}
\usepackage{xcolor}
\definecolor{CX}{rgb}{0, 128, 128}

\usepackage[bookmarks,backref=true,linkcolor=black]{hyperref} 
\hypersetup{
 pdfauthor = {},
 pdftitle = {},
 pdfsubject = {},
 pdfkeywords = {},
 colorlinks=true,
 linkcolor= black,
 citecolor= black,
 pageanchor=true,
 urlcolor = black,
 plainpages = false,
 linktocpage
}

\onlineid{1382} 
\vgtccategory{Theoretical \& Empirical} 
\vgtcinsertpkg 

\title{What Does the Chart Say? Grouping Cues Guide Viewer Comparisons and Conclusions in Bar Charts}

\author{Cindy Xiong Bearfield, Chase Stokes, Andrew Lovett, and Steven Franconeri}
\authorfooter{ 
\item
Cindy Xiong Bearfield is with Georgia Tech. 
\item
Chase Stokes is with University of California, Berkeley.
\item
Andrew Lovett is with the US Naval Research Lab.
\item
Steven Franconeri is with Northwestern University.
}

\shortauthortitle{Xiong Bearfield \MakeLowercase{\textit{et al.}}:What Does the Graph Say: Grouping Cues Guide Viewer Conclusions in Bar Charts}

\abstract{Reading a visualization is like reading a paragraph. Each sentence is a comparison: the mean of these is higher than those; this difference is smaller than that. What determines which comparisons are made first? The viewer's goals and expertise matter, but the way that values are visually grouped together within the chart also impacts those comparisons. Research from psychology suggests that comparisons involve multiple steps. First, the viewer divides the visualization into a set of units. This might include a single bar or a grouped set of bars. Then the viewer selects and compares two of these units, perhaps noting that one pair of bars is longer than another. Viewers might take an additional third step and perform a second-order comparison, perhaps determining that the difference between one pair of bars is greater than the difference between another pair. We create a visual comparison taxonomy that allows us to develop and test a sequence of hypotheses about which comparisons people are more likely to make when reading a visualization. We find that people tend to compare two groups before comparing two individual bars and that second-order comparisons are rare. Visual cues like spatial proximity and color can influence which elements are grouped together and selected for comparison, with spatial proximity being a stronger grouping cue. Interestingly, once the viewer grouped together and compared a set of bars, regardless of whether the group is formed by spatial proximity or color similarity, they no longer consider other possible groupings in their comparisons.
} 


\begin{document}
\maketitle
\section{Introduction}
Comparison is a foundational perceptual operation in data visualizations \cite{gleicher2011visual}. Even for an expert, reading a chart is less like instantly recognizing a picture and more like slowly reading a paragraph \cite{shah2011bar}, with each sentence a pattern one could see.
Some of the most important sentences are individual comparisons that unfold as a sequence over time \cite{nothelfer2019measures}. 
Some comparisons may be obvious, appearing early in the sequence for most viewers. Other comparisons may appear later, or not at all. 
In this way, comparisons are situated in chart comprehension as a fundamental part of how readers extract information from a chart - performing elementary tasks to determine the basic values, intermediate tasks to find relationships between these values, and overall comprehension tasks that bring in external context or domain-specific knowledge to inform the interpretation of these comparisons \cite{friel2001making}. 
Thus, this paper focuses on investigating how visualizations can be designed to guide the viewer toward making the appropriate comparisons so that they can access the information the visualization was meant to convey.

\begin{figure}[!ht]
 \includegraphics[width=3.4in]{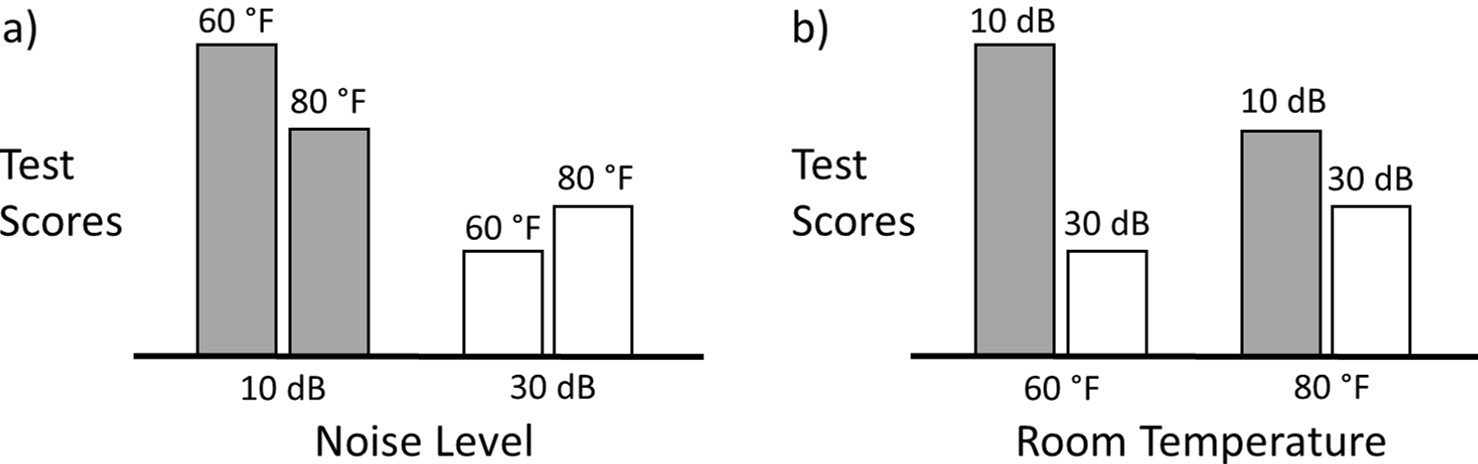}
 \caption{Bar charts showing how noise level and room temperature influence test scores. Adapted from \cite{shah2002review}.}
 \label{fig:intro}
\end{figure}

While top-down processing mechanisms such as the viewers' goals and expertise can influence which comparisons are made, comparisons can be;'/ profoundly impacted by how visual elements are arranged, including their location, color, and size. For example, consider the two bar charts in Figure \ref{fig:intro}. Although the charts reflect an identical underlying dataset, each chart's layout emphasizes different comparisons. In the left chart, readers may be inclined to compare noise levels. For example, concluding that test scores are higher when the noise level is 10 dB. In contrast, when viewing the right chart, they may be inclined to compare temperatures, instead concluding that test scores are higher when the temperature is 60$^{\circ}$, provided the noise level is also 10 dB \cite{shah2002review}. 
The affordances of the two charts make certain comparisons (and therefore conclusions) more intuitive for readers.

Figures like Figure \ref{fig:intro} are commonplace visualizations that appear in scientific journals and news articles. Researchers and analysts rely on them to communicate information, and viewers rely on them to make decisions, such as whether the results are valid or interesting, and how this might influence their future behavior. There are many ways to arrange even this simple data set. For example, in addition to the two arrangements shown in Figure \ref{fig:intro}, one could keep the noise-level grouping in a) and alternate the colors, or keep the temperature grouping in b) and color all the bars the same. Considering the combinatorial explosions for both the number of comparisons one could make for a visualization and the number of ways the data could be arranged, it is critical that a visualization leads the viewer to make the comparison that helps them see the `right' story.

\textbf{Contributions:} Building on established psychology and cognitive science research, we developed a taxonomy (e.g., see Table \ref{comparisonTypes}) of comparisons and designed two empirical studies examining the role of bottom-up attention guided by visual grouping cues such as spatial proximity and color in visual data comparison tasks. 
We collected and categorized hundreds of verbal descriptions of visual comparisons made in simple bar charts that varied systematically in the way that their data marks were visually grouped. We present both informal guidelines for designers and formal rules that can be implemented in recommendation systems for improving the efficiency, accuracy, and accessibility of data communication.

\section{Related Work}
\label{relatedWork}

\textbf{Comparison} is a core data visualization task \cite{amar2005low, gleicher2017considerations, gaba2022comparison}. One famous early example by William Playfair was intended to facilitate a comparison between import and export routes between England and the East Indies \cite{cleveland1984graphical}. Comparison requires picking out a mark or set of marks, computing the magnitude of some value or statistic across them, and noting the magnitude's polarity (ordinal comparison) and/or size (metric comparison) compared to another. For humans, making comparisons entails at least two steps: grouping and actual comparison. 

In the first step, the viewer \textbf{visually groups} the visualization components into marks or sets of marks, defining the units to be compared \cite{pinker1990theory}. For example, when viewing the bar chart in Figure \ref{fig:intro}a, one might group the two bars depicting a noise level of 10 dB on the left together, and group the two bars depicting 30 dB on the right together. This grouping process has a long history of study in perceptual psychology \cite{brooks2015traditional}. People can choose to group an arbitrary set of objects by jointly attending to them \cite{egeth2010salience}, but that operation is tough and capacity-limited \cite{franconeri2007many}. Groupings are more often processed as default visual operations that are commonly used across many types of images. These defaults include grouping objects by `spatial' and `featural' cues. Spatial cues include proximity (combining objects that are near each other into a single unit), connectedness (combining objects that are visibly connected), co-linearity (combining objects that lie along a common line), and region-sharing (combining objects that lie within the same region). 
Featural cues include color, shapes, or sizes. When grouping cues compete, spatial cues tend to beat featural cues and are harder to inhibit, even when they are counterproductive to effective task performance \cite{brooks2015traditional, wagemans2012century}. Spatial cues also appear to be processed in parallel across a display \cite{franconeri2009number, theeuwes1996parallel}, in contrast to featural cues which are argued to be grouped by only one value (a single color, shape, or size) at a time \cite{huang2002symmetry, yu2019similarity, yu2019gestalt, huang2007boolean}. 
People can enumerate up to three groups of elements sharing the same color in a set with many overlapping items \cite{halberda2006multiple}.
Spatially proximate items can also be grouped when they are visually dissimilar (e.g., two adjacent dots with different colors) \cite{wagemans2012century}, and visually similar features (e.g., two dots with the same color) can be grouped even when they are spatially separated \cite{wolfe2004attributes}. 
Spatial proximity and color can both act as grouping cues by pushing/constraining participants to perform mental arithmetic on sets of dots \cite{ciccione2020grouping, xiong2022investigating}.

Among featural grouping cues, research suggests that some grouping features lead to better performance in distinguishing groups of marks. In multi-class scatterplots, groups with different colors are more accurately distinguished than groups with different shapes \cite{duncan1989visual, gleicher2013perception}. Relying on perceptually-modulated color spaces can help a designer choose sets of colors that are maximally distinguishable, thus forming groupings that are most easily dissociated from other groups \cite{brewer2003colorbrewer}. Taking inspiration from older perceptual psychology work \cite{hegde1999popout}, similar work is beginning to appear for shape spaces with the goal of designing sets of marks that are maximally mutually distinguishable \cite{burlinson2017open, huang2020space, demiralp2014learning}. Motion cues like common fate \cite{levinthal2011common} also automatically lead to grouping in visualizations, and recent work compares its strength to featural cues in visualization contexts \cite{chalbi2019common}.

In the second step, two units or grouped units are selected, and they are \textbf{compared} to extract some difference or similarity between them \cite{shah2011bar, michal2017visual, michal2016visual}. For example, when viewing the bar chart in Figure \ref{fig:intro}a, one might compare the two leftmost bars and conclude ``When the noise level is 10 dB, 60$^{\circ}${} is better than 80$^{\circ}$.'' Alternatively, after spatially grouping the two leftmost bars together and grouping the two rightmost bars together, one might compare the two groups and conclude ``10 dB is better than 30 dB.'' In fact, there is a combinatorial explosion of possible comparisons that might be made, depending on how the chart objects are visually grouped in the first step, and which units are selected for comparison in the second step. Even a simple 2-bar graph enables at least a dozen different comparisons \cite{michal2017visual}.

The present experiments purposely strip most context and meaning from the tested visualizations, in an attempt to isolate the influence of spatial and color grouping cues that should be common across any visualization. There are of course many other factors that will influence what people compare in a visualization, including their goals, experience, annotations, and other context. But the perceptual psychology literature shows that there is a unique contribution of spatial and featural grouping cues \cite{lamy2004effects}, regardless of these factors, and it is thus critical to model their influence.

\begin{table*}[!ht]
\centering
\rowcolors{2}{gray!25}{white}
    \scriptsize
    \caption{Comparison types with concrete examples.}. 
    \begin{tabular}{lp{1.5cm}p{7.65cm}ll}
   \textbf{ } & \textbf{Visual} & \textbf{Example Description} & \textbf{Category} & \textbf{Detail} \\ 
   \hline 
    1 & \parbox[c]{1em}{\includegraphics[height = 1.3cm]{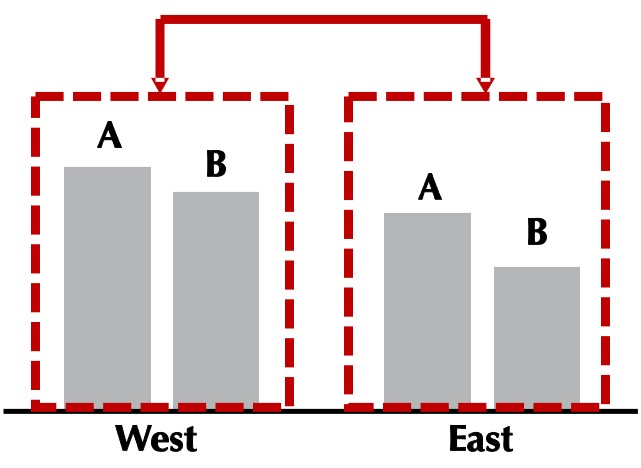}}
    & East is worse than West.  & Group Comparison & Near \\ 
    
    2 & \parbox[c]{1em}{\includegraphics[height = 1.3cm]{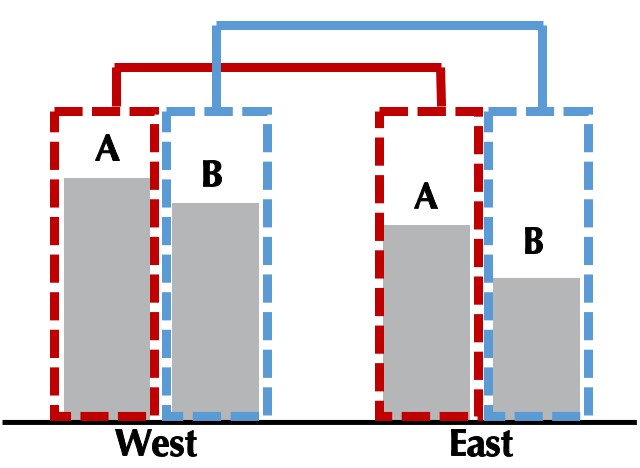}}
    & A has more sales than B.  & Group Comparison & Far \\ 
    
    3 & \parbox[c]{1em}{\includegraphics[height = 1.3cm]{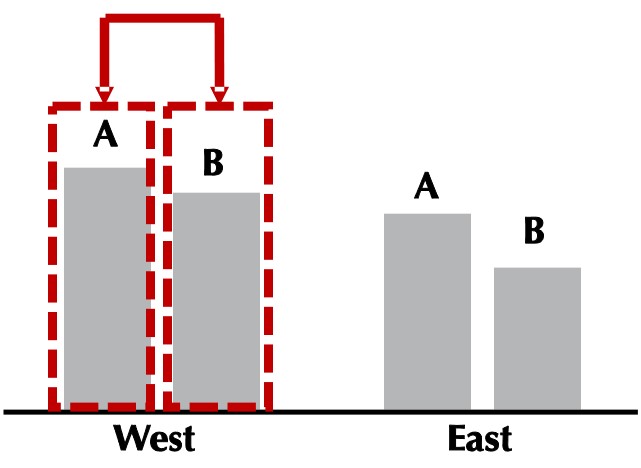}}
    & A's West region generates more revenue than that of B. & Pair Comparison & Near \\ 
    
    4 & \parbox[c]{1em}{\includegraphics[height = 1.3cm]{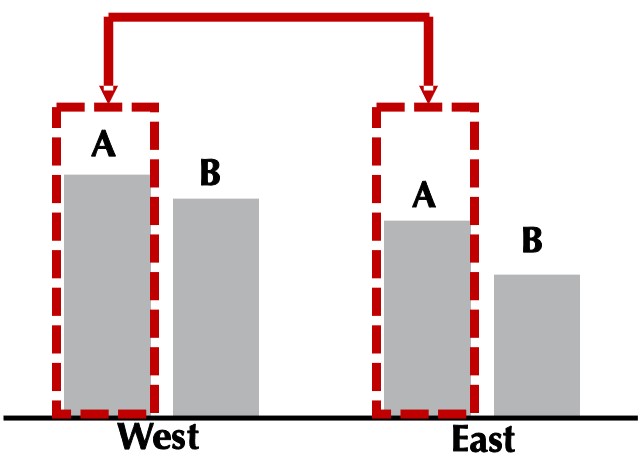}}
    & A does much better in the West as opposed to the East.  & Pair Comparison & Far \\ 
    
    5 & \parbox[c]{1em}{\includegraphics[height = 1.3cm]{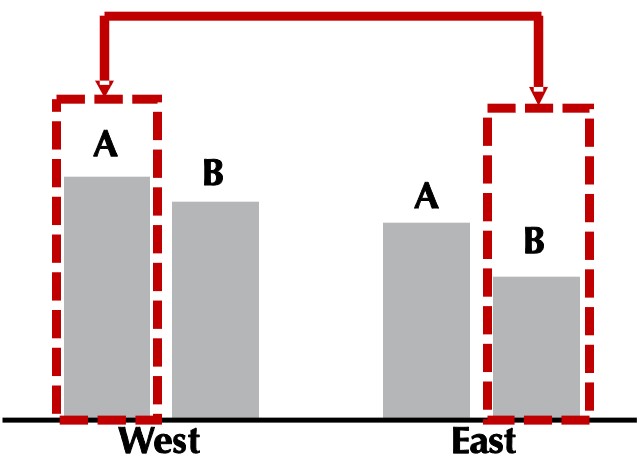}}
    & A's West region has higher revenue than B's East region.  & Pair Comparison & Edge \\ 
    
    6 & \parbox[c]{1em}{\includegraphics[height = 1.3cm]{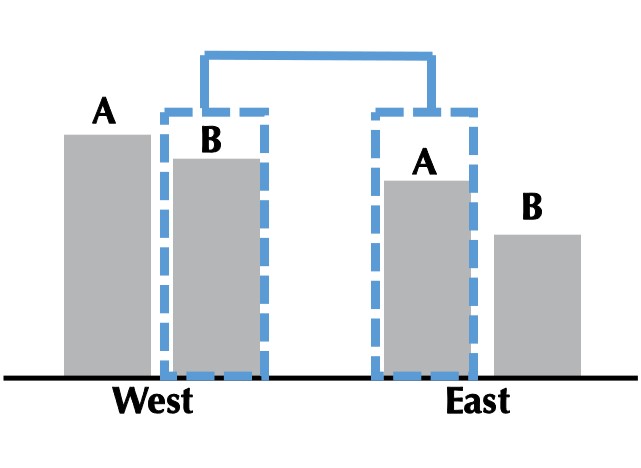}}
    & B's West region has higher revenue than A's East region. & Pair Comparison & Center \\ 
    
    7 & \parbox[c]{1em}{\includegraphics[height = 1.3cm]{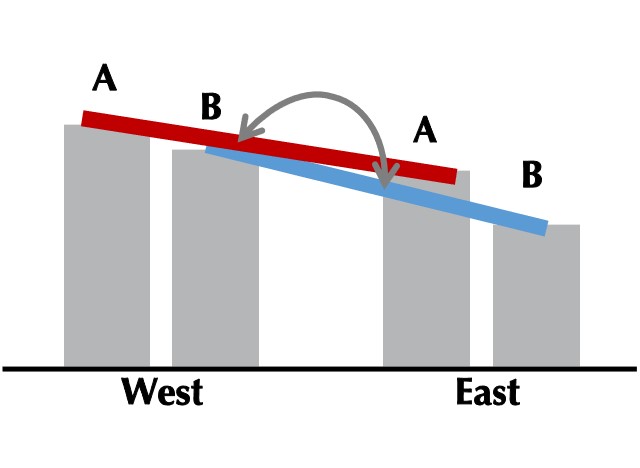}}
    & A shows more difference in revenue between the two regions than B. & 2\textsuperscript{nd} Order Quantity Comparison & Far-First\\ 
    
    8 & \parbox[c]{1em}{\includegraphics[height = 1.3cm]{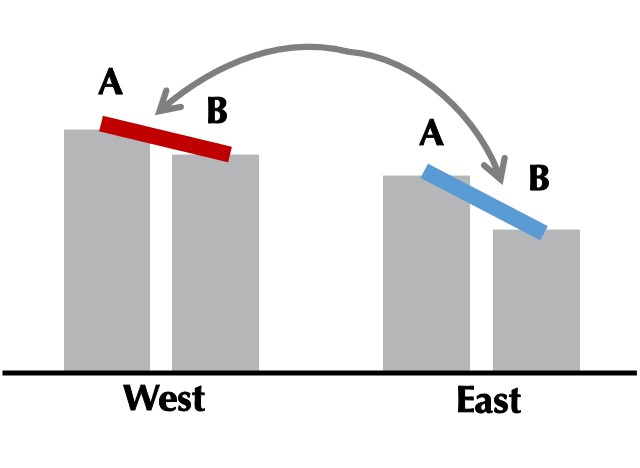}}
    & West's profitability fluctuates more than East. & 2\textsuperscript{nd} Order Quantity Comparison & Near-First\\ 
    
    9 & \parbox[c]{1em}{\includegraphics[height = 1.3cm]{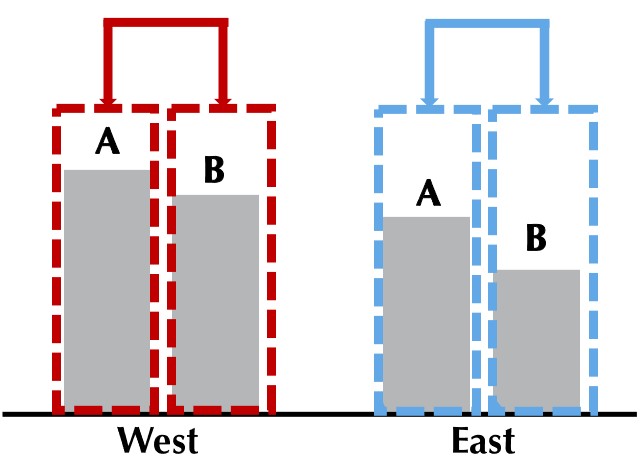}}
    & West A is greater than West B \textit{and} East A is greater than East B. & 2\textsuperscript{nd} Order Relation Comparison & Near-First\\ 
    
    10 & \parbox[c]{1em}{\includegraphics[height = 1.3cm]{Figure/main_distal.jpg}}
    & West A is greater than East A \textit{and} West B is greater than East B. & 2\textsuperscript{nd} Order Relation Comparison & Far-First\\ 
    
    11 & \parbox[c]{1em}{\includegraphics[height = 1.3cm]{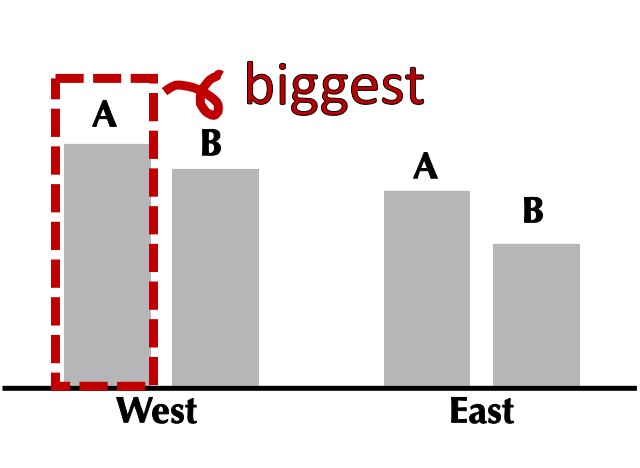}}
    & Company A West is the best. & Superlative &  Maximum \\
\hline
\end{tabular}
\label{comparisonTypes}
\end{table*}


\section{Theoretical Framework and Hypotheses}

\label{types}
We recognize that there are few established taxonomies categorizing the type of comparisons a viewer might make upon viewing a particular visualization, and even fewer that examine the cognitive and perceptual mechanisms behind visual comparison tasks in visualizations \cite{gleicher2017considerations, xiong2021visual}. In this paper, we take an initial step at mapping out a comparison taxonomy for simple bar visualizations, referencing relevant existing work in psychology, cognitive science, and research methods \cite{cozby2007methods}. We use a 2x2 bar chart (similar to the ones shown in Figure \ref{fig:intro}) illustrating the revenue of two companies (A and B) in two regions (East and West) as a running example to showcase the various types of comparisons people might make (see Table \ref{comparisonTypes}). In the later sections, we empirically demonstrate that this taxonomy matches what viewers do upon seeing a visualization via a pilot experiment. Using this taxonomy, we formulate hypotheses predicting how the spatial, and color grouping of a visualization could affect viewer comparison behaviors. 

When classifying comparisons, we must first consider how elements have been grouped together, since grouping determines the units that are available to be compared. Typically, the visual system provides ``the forest before the trees,'' allowing people to easily see groups of objects, whereas more time and effort is required to perceive the individual objects within those groups \cite{ahissar2004reverse,navon1977forest,love1999structural}. Notably, there is evidence that this preference for larger-scale groups is culture-specific \cite{lovett2011cultural}, and even within a culture it may be possible to prime individuals to focus on smaller-scale details \cite{xiong2019curse}. However, in typical visualization-viewing contexts, we can predict that the easiest comparisons will be \emph{group} comparisons, where pairs of bars are grouped together and then compared (Hypothesis 1). In Table \ref{comparisonTypes}, we distinguish between \emph{near} groups, where neighboring bars are grouped together, and \emph{far} groups, where most distant bars are grouped together. Because spatial proximity is a strong grouping cue, we can predict that near groups will be more common.

The second step in classifying a comparison is determining which units have been selected for comparison. For the near group and far group categories, this step is trivial--once pairs of bars have been grouped together, there are only two groups available for comparison, so those will be selected.

Thanks to the flexible nature of the mind, perceptually grouped units are not the only things that can be compared. It is also possible to perform a comparison over the results of other comparisons, a process termed \emph{second-order comparison} \cite{lovett2009solving,lovett2012modeling}. For example (Table \ref{comparisonTypes}, line 8), one might compare the two leftmost bars to get the difference between companies A and B in the West, then compare the rightmost bars to get the difference between companies A and B in the East, and finally compare these two differences to determine that the difference between companies is greater in the West than in the East. This example illustrates a second-order \emph{quantity} comparison because one quantitative difference is being compared to another quantitative difference to determine which is greater. We can predict that these comparisons will be rarer than group or pair comparisons because they require more effort: instead of performing a single comparison operation, individuals must perform three comparison operations (Hypothesis 3). In Table \ref{comparisonTypes}, we note that there are two types of second-order quantitative comparisons: near-first, which begins with comparing two near bars, and far-first, which begins with comparing two far bars.

There is a second type of second-order comparison that is more integrated with perceptual grouping. Suppose an individual begins by generating near groups. Rather than comparing the groups, they might compare the two bars within each group to produce a \emph{structured representation} for that group. A structured representation describes relationships between elements within a unit \cite{gentner1983structure,biederman1987}. For example, in Table \ref{comparisonTypes}, line 9, a structured representation of the left bars would indicate that A is greater than B. Similarly, a structured representation of the right bars would also indicate that A is greater than B. These two structured comparisons could then be compared to determine that A makes more money than B in both the West and the East. Note that this comparison, which we term second-order \emph{relation} comparison, is distinct from second-order quantity comparison where two quantitative differences are being compared. Here, a simple ``greater than'' relationship is being compared, to determine whether A is greater than B in both the West and the East. This distinction is important because a large body of research suggests humans are adept at comparing structured representations to find commonalities or differences in their relations \cite{gentner1983structure,markman1996commonalities,hummel1997distributed}. Second-order relation comparisons should be more difficult and thus rarer than group or pair comparisons because they require three comparison operations. However, we predict that they will be more common than second-order \emph{quantity} comparisons because humans are skilled at performing relational comparisons over structured representations (Hypothesis 3).

Comparisons are not just `-er' (e.g., higher, better), they can be `-est' too (e.g., highest, best), such as identifying the minimum and maximum points in a chart, as shown in row 11 of Table \ref{comparisonTypes}. Unlike comparative relations (e.g., X is bigger/taller/longer than Y), which take more effort to extract because they require shifting attention between two elements \cite{michal2017visual, wolfe1998can}, finding the maximal or minimal element is automatic and effortless \cite{picon2017finding}. Thus we hypothesize that these superlative comparisons will be both the first and most frequent comparison that viewers make (Hypothesis 5).

Despite the strong bias towards grouping elements together, viewers are also able to perceive the individual bars in the bar chart. If the viewer focuses on the individual bars, then they can select any two of the four total bars for comparison. We term such comparisons \emph{pair} comparisons. There are four possible pair comparisons, based on which two bars are selected: \emph{near} comparisons, between two adjacent bars (similar to near groups, where two adjacent bars are grouped together); \emph{far} comparisons, between the corresponding bars in the two groups, for example the two A bars; \emph{edge} comparisons, between the leftmost and rightmost bars; and \emph{center} comparisons, between the two innermost bars. We view the near and far comparisons as the more useful, as they allow the viewer to compare either two companies within a region or two regions within a company (in our running example), so we predict these comparisons to be made more often (Hypothesis 2). From among near and far comparisons, we predict near comparisons will be selected more often, just as near groups should be preferred over far groups, because once the visual system selects one element as the focus of attention, it typically becomes more sensitive to nearby elements \cite{posner1980,folk2010}, and thus more likely to select a nearby element for comparison. Similarly, just as individuals might be more likely to \emph{group} bars that share features like color and size (Hypothesis 6), they may be more likely to \emph{compare} visually similar pairs (Hypothesis 7).

One final point must be considered for group vs. pair comparisons. We have suggested that group comparisons will often be preferred because the visual system automatically organizes visual input into groups. However, not all groups are equally available. For example, if spatially proximate bars are grouped together, this supports a near-group comparison but not a far-group comparison. To perform a far group comparison, a viewer would need to abandon the initial grouping, shift to reasoning over the individual bars (which takes effort), and then group the bars together in a new way (which also takes effort). An analysis of a popular visual intelligence test indicated that this process, known as \emph{perceptual reorganization}, is required by some of the most difficult problems on the test \cite{lovett2017modeling}, suggesting that individuals would rarely perform such a process while viewing a chart. Thus, we can see why grouping cues are critically important: if viewers group the wrong chart elements together, they may never make the comparison that will produce the desired conclusion. In the present work, the difficulty of perceptual reorganization means that after individuals perform a near-group comparison, they should be less likely to perform a far-group comparison, and vice versa (Hypothesis 8).

\begin{figure*}[h!]
 \includegraphics[width=\linewidth]{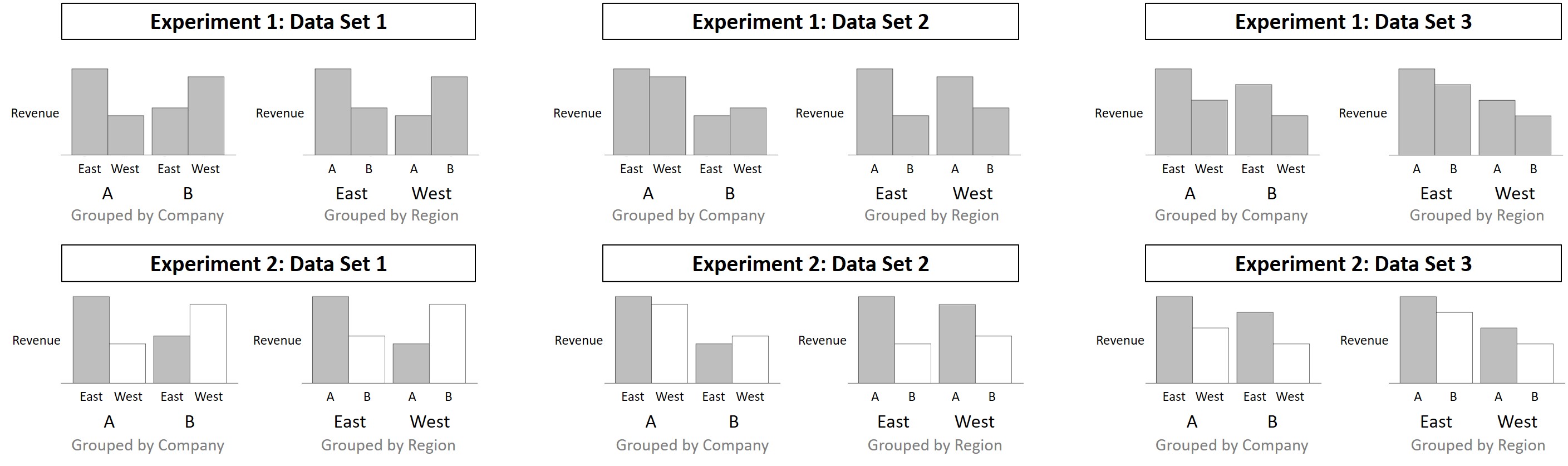}
 \caption{Stimuli for Experiments 1 and 2.}
 \label{stimulus}
\end{figure*}

\subsection{Experimental Hypotheses}
We summarize the eight hypotheses motivated in the previous section below:
\begin{enumerate} [noitemsep]
  \item Viewers will make group comparisons before making pair comparisons. 
  \item Among pair comparisons, viewers will make more near and far pair comparisons and fewer edge and center pair comparisons.
  \item Second-order comparisons (both quantity and relation) will be made less frequently than group comparisons and pair comparisons. Among second-order comparisons, relation comparisons will be more common than quantity comparisons. 
  \item Superlative comparisons will be made the earliest and most frequently.
  \item Among group comparisons, viewers will more often group spatially proximate bars, but this tendency will be moderated by the color and size similarities of the bars. 
  \item Among pair comparisons, viewers will more often compare spatially proximate bars, but this tendency will be moderated by the color and size similarities of the bars.
  \item Among group comparisons, once people make a near-group comparison, they will be less likely to make a far-group comparison, and vice versa. 
\end{enumerate}

\section{Pilot Experiment}

We conducted a crowdsourced pilot study with two goals in mind: first, to test whether crowdworkers on Amazon's Mechanical Turk (MTurk) could produce verbal sentence responses that we could interpret to confidently assess which comparison type they made, and second, to check whether our taxonomy based in existing psychology and cognitive science literature covers the comparisons that people make. To do this, we recruited 58 participants on MTurk. In a Qualtrics survey \cite{snow2013qualtrics}, participants viewed a bar chart depicting the revenues of two companies (A and B) in two regions (East and West), similar to the ones shown in Table \ref{comparisonTypes}. 

Participants were prompted to type the first conclusion they drew from the chart. After they typed their first conclusion, a new prompt appeared on the same page and asked them to enter a second conclusion. This continued until the participant entered five conclusions. Then, on a separate page, participants entered demographic information such as their age and level of education. The order in which they wrote the conclusions was recorded as the sentence \textit{rankings}. 

All responses were interpreted by two human coders and categorized following the taxonomy in Table \ref{comparisonTypes}. While detailed results from the pilot can be found in the supplementary materials, we discuss some insights we gained from the pilot that helped us finalize our experimental designs. We discuss the potential of using this coding system for creating recommendation systems in Section \ref{guidelines}.

\subsection{Insights from the Pilot}
\label{insightFromPilot}
We noticed that several participants mentioned multiple comparisons in the same conclusion. For example, when prompted to report the first conclusion, they identified both a group comparison and a pair comparison. To account for both comparison types made, we broke up the response in two parts and included each in its corresponding category, preserving the rank order for both. We also noticed that many participants stopped giving interpretable responses after the third conclusion, typing responses such as ``I can't draw any additional conclusions.'' We, therefore, limited the total number of comparison prompts to three for our experiments.

Several participants in the pilot also gave responses that did not fall in the categories defined in Table \ref{comparisonTypes}. These all included comments about the number of companies shown, the colors used in the visualization, or axis labels. We decided to include an additional category of \textit{Other} to capture these conclusions. Furthermore, some participants commented on the implications, novelty, or ``obviousness'' of data, such as saying ``I would invest in company A in the future.'' Following the approach taken in \cite{shah2011bar} for similar responses, we included conclusions like these in the \textit{Other} category as well.

Additionally, we observed that very few conclusions were second-order quantitative or relation comparisons. This may reflect second-order comparisons being cognitively more difficult to extract, but it also could be due to the particular set of data values we used in the pilot. For our experiments, we, therefore, decided to create three distinct data sets, illustrated in Figure \ref{stimulus}. This approach not only helps us generalize our findings across different data values but also sheds light on whether the underlying relationship between the data values shown in the bar chart, in order words, the size of the bars, would affect the types of comparisons viewers make.

For our experiments, we continued the theme of showing the revenue of two companies (A, B) across two regions (East, West). As shown in Figure \ref{stimulus}, we kept the total value of the four data points identical across our three data sets, but we varied which data value represents the revenue for which company and region. For example, the smallest bars in data set 1 and data set 2 are identical in height. But in data set 1, the smallest bar represents the revenue of West A, whereas in dataset 2, this smallest bar represents the revenue of East B. 

In data set 1, if a viewer groups the two companies together, they will realize that the average revenue of the two companies is similar. Similarly, if they group the two regions together, they will see that the average revenue across the two regions is similar. However, there is a drastic difference within each company, such that one region makes significantly more than the other region. In data set 2, overall company A makes significantly more revenue than company B, and within each company, their two regions make similar amounts of revenue. In data set 3, overall company A makes more than company B, and overall the East region makes more than the West region.



\begin{figure*}[h!]
 \includegraphics[width=\linewidth]{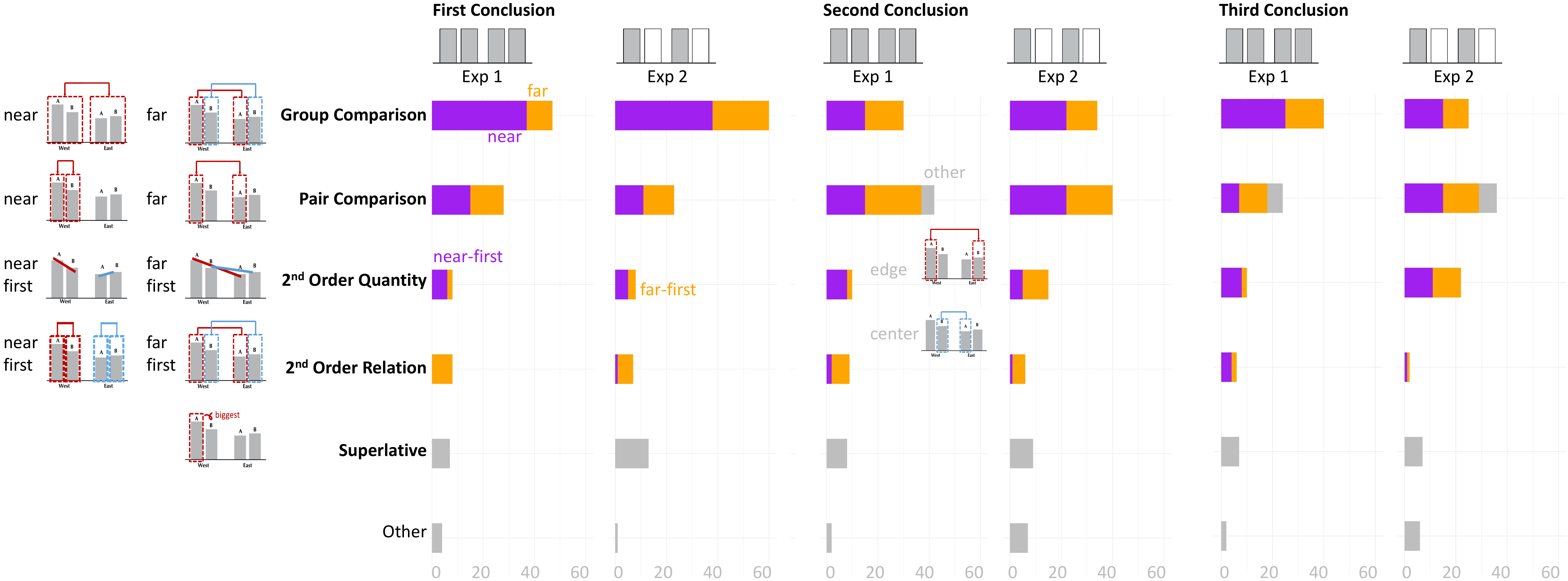}
 \caption{Total number of comparisons of each type in Experiments 1 and 2. The first comparison participants make upon seeing a 2x2 bar chart is most likely a near-group comparison.}
 \label{exp12ComparisonOverview}
\end{figure*}

\section{Experiment 1}

Experiment 1 allowed us to evaluate Hypotheses 1-5, which concern how frequently and how early the different comparison types will be performed. A power analysis based on the pilot effect sizes derived from logistic regression models suggested a target sample of 45 conclusions per chart would give 80\% power to detect patterns in conclusion frequencies at an alpha level of 0.05 \cite{hsieh1998simple}. Because we have no control over the conclusions a given participant would draw from a visualization, we iteratively surveyed and excluded participants on Amazon Mechanical Turk who gave nonsensical answers or failed attention tests (e.g., `How many years of experience of Computer Programming do you have? Please choose ``10+''  to show that you are paying attention.') until we had at least 45 conclusions for each bar chart.

\subsection{Participant}
We collected sentence conclusions from 104 unique participants ($M_{age} = 39.86$, $SD_{age} = 9.89$, 40 females), each providing three conclusions in the order of which came to mind first. After excluding the nonsensical conclusions (7.6\%), we ended up with 292 conclusions. The participants were compensated at a rate of 9 USD per hour. 

\subsection{Method and Procedure}
This experiment followed the same procedure as the pilot experiment. 
Participants read bar charts and were prompted with, ``Please write down the FIRST/SECOND/THIRD conclusion you drew from this data.'' 
In addition to preparing three data sets, we manipulated the spatial grouping of each data set, such that half of them were spatially grouped by company (A, B), and the other half spatially grouped by region (East, West), to create a total of six charts, as shown in Figure \ref{stimulus}. 
We adopted a between-subject experimental design, where each participant was randomly assigned to view one of the six bar charts. After viewing the bar chart, the participants were prompted to write down their top \textit{three} conclusions.

Two coders coded each conclusion into one of the eleven comparison types listed in Table \ref{comparisonTypes}. Discrepancies between coder ratings were resolved through discussions. Overall, the two coders agreed 83.2\% of the time in their ratings, with a high inter-rater reliability Kappa value of 0.798 ($z = 32.9$, $p < 0.001$).



\subsection{Results Overview}

As shown in Figure \ref{exp12ComparisonOverview}, for the first conclusion, participants made group comparisons most often (46.1\%), followed by pair comparisons (27.5\%), both of which were significantly more frequent than the other comparison types (${\chi}^2$ = 85.41, $p < 0.001$). For the second conclusion, participants made pair comparisons most often (41.6\%), followed by group comparisons (29.7\%), and again both significantly more often than the other comparisons types (${\chi}^2$ = 72.05, $p < 0.001$). For the third conclusion, similar to the first conclusion, participants made more group comparisons (44.9\%), followed by pair comparisons (27.0\%), both of which were made significantly more often than the other comparison types (${\chi}^2$ = 70.43, $p < 0.001$).

\begin{figure*}[!ht]
 \includegraphics[width=\linewidth]{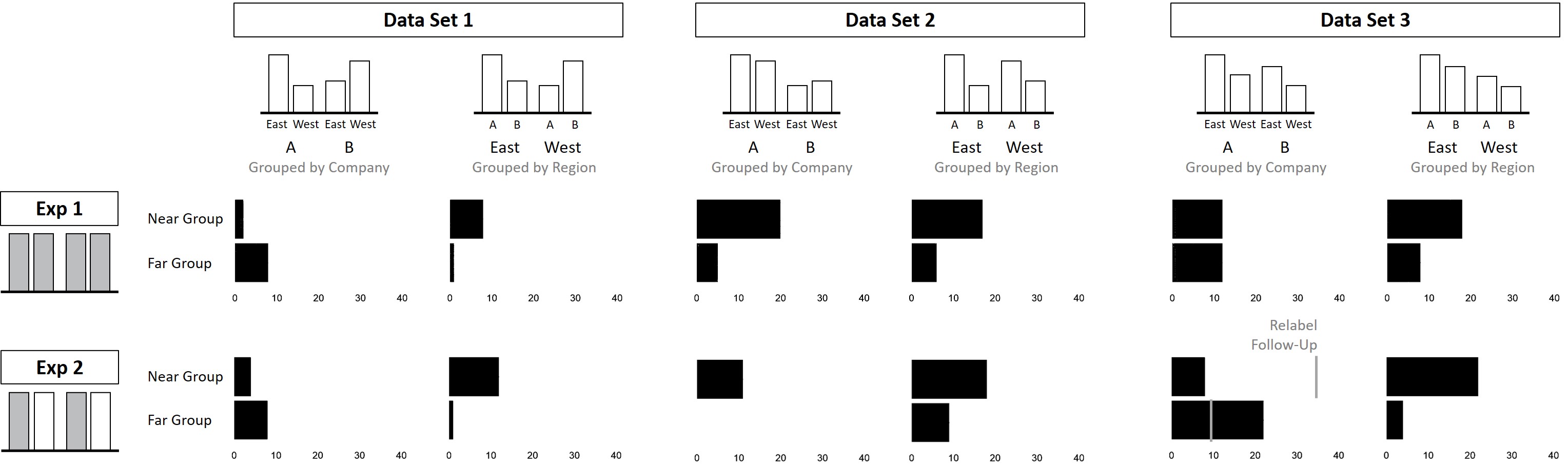}
 \caption{Total number of near and far \textbf{group} comparisons made while viewing the six different charts in Experiments 1 and 2. Participants generally grouped by spatial proximity, leading to more frequent near-group comparisons (exceptions of Data Set 1 by Company and Data Set 3 by Company). In comparing the experiments, color did not have a strong impact on grouping in most cases (except for Data Set 3 by Company). The Relabel Follow-Up indicated an effect of labels on group comparisons. This is further discussed in the results.
 }
 \label{exp12groupComparisons}
\end{figure*}

\subsection{Hypothesis 1: Group vs. Pair Comparisons}

To investigate whether participants made group comparisons before making pair comparisons, we isolated the two comparison types for further investigation. We conducted a multinomial logistic regression, using the \textit{nnet} library in R \cite{ripley2016package}, to predict the likelihood of the order in which a viewer would conduct a group or a pair comparisons. We found that participants were the most likely to make group comparisons first and pair comparisons second. More specifically, they were 2.35 times more likely to make a pair comparison as their second conclusion compared to their first conclusion (p = 0.011). They were equally not likely to make pair comparisons as their first and third conclusions (odds ratio = 1.01, p = 0.98). 

For group comparisons, by the second conclusion, the participants were less than half as likely to make a group comparisons as their second conclusion (odds ratio = 0.425, p = 0.011). However, they were equally as likely to make group comparisons as their first and third conclusions (odds ratio = 0.99, p = 0.98). This supports our hypothesis that viewers tend to make group comparisons before making pair comparisons.

\subsection{Hypothesis 2: Types of Pair Comparisons}

We counted the total number of center, edge, far, and near type pair comparisons, and found there to be significantly more far (48.9\%, Bonferroni-corrected p-value = 0.002) and near (39.4\%, $p < 0.001$) pair comparisons, compared to center (9.6\%) and edge (2.1\%) pair comparisons (${\chi}^2$ = 70.43, $p < 0.001$), as we hypothesized. Interestingly, the number of far and near pair comparisons did not differ significantly. We further consider far/near pair comparisons when testing Hypothesis 7 in Experiment 2. 




\subsection{Hypothesis 3: Second Order Comparisons}

From Figure \ref{exp12ComparisonOverview} we can see that second-order comparisons, both quantity and relation types, are less common compared to group and pair comparisons (${\chi}^2$ = 70.43, $p < 0.001$). This supports our Hypothesis 3. However, we did not find a significant difference between the likelihood of participants making second-order quantity or relation comparisons, across all three rank orders (${\chi}^2$ = 0.57, $p = 0.75$). Still, because the total number of second-order comparisons made is small, there could have been differences between the frequency with which viewers make second-order quantity and relation comparisons that we were unable to detect. We discuss this limitation in Section \ref{limitation} and point out future directions to further explore this hypothesis. 

\subsection{Hypothesis 4: Superlatives}

Similarly, from Figure \ref{exp12ComparisonOverview}, we see that, contrary to our hypothesis, people did not make superlative comparisons the most frequently, despite research in human perception suggesting that finding maximums and minimums tend to be less effortful. We suspect this might be one scenario where rules that apply to how humans perceive the natural world failed to account for human behavior while interacting with visualizations. When participants are asked to draw conclusions from a bar chart, they might have a  `complexity threshold' for what type of comparisons would qualify as a valid conclusion. The simplicity of noticing a maximum or a minimum value might not fall past this threshold for most people. Alternatively, it is also possible that extracting minima and maxima from a chart may be less preferred when they are only a few data points, or that the superlative outlier needs to be significantly different from the rest of the data points. We further discuss the implication of this finding in Section \ref{limitation}. 

Additionally, there is no significant difference between the order in which participants made superlative comparisons. They were equally likely to make superlative comparisons as their first, second, or third conclusion (${\chi}^2$ = 0.09, $p = 0.96$).

\begin{figure*}[!ht]
 \includegraphics[width=\linewidth]{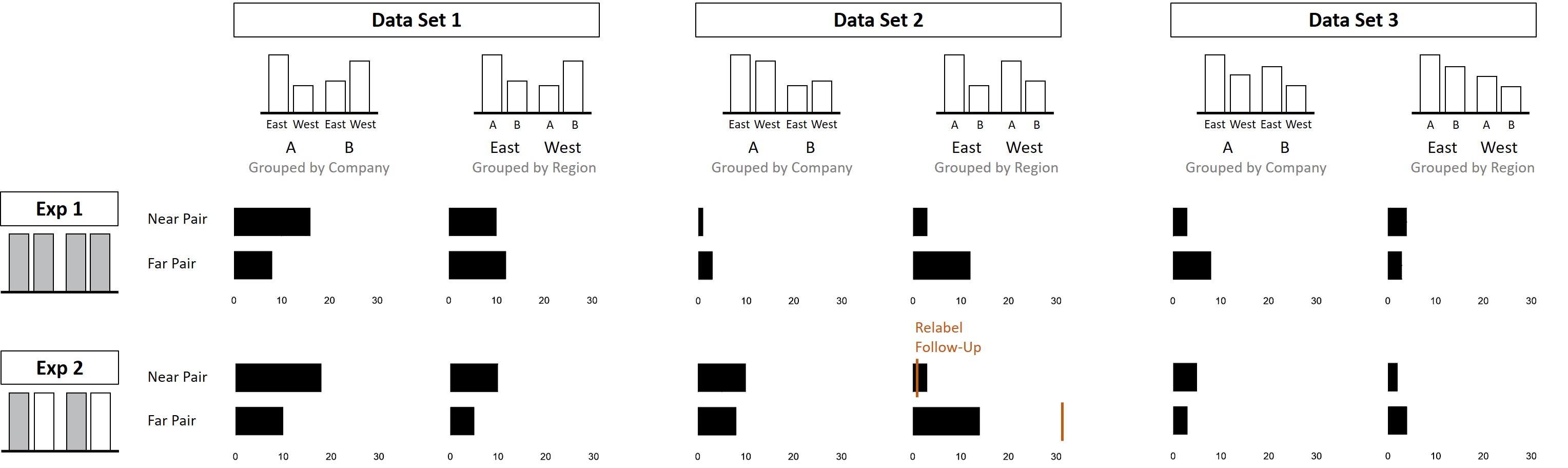}
 \caption{Total number of near and far \textbf{pair} comparisons made while viewing the six different charts in Experiments 1 and 2. Overall, participants made near and far pair comparisons at similar frequencies (exception of Data Set 2). In comparing the experiments, color did not have a strong impact on pair comparisons in most cases. The Relabel Follow-Up also did not indicate a strong effect of labels on pair comparisons.}
 \label{exp12pairComparison}
\end{figure*}

\section{Experiment 2}

Experiment 2 allowed us to evaluate Hypotheses 6, 7, and 8, which concern the effect of visual grouping cues on the comparisons that viewers make. The visual grouping cues we investigated were spatial proximity and color, but the inclusion of the different underlying data values also allowed us to gain some insights into how the bar sizes interact with spatial proximity and color to impact comparisons. 

Spatial proximity describes the distance between pairs of bars. For example, in Figure \ref{stimulus}'s Data Set 1, the Grouped by Company chart places the two bars for a company beside each other, whereas the Grouped By Region chart places the two bars for a region beside each other. In Experiment 1, there is no color grouping cue because all the bars have the same color. We recolored the charts from Experiment 1 to alternate between grey and white, as shown at the bottom of Figure \ref{stimulus}, to examine the effect of color cues. Under this color scheme, the two bars that are beside each other have different colors. Thus, we can pit the effect of spatial grouping cues against color grouping cues, to see which cue has a stronger impact on what people compare. 

The size of the bars is a special case because it is determined by the underlying values of the data: if two bars happen to represent data with similar values, those bars will have similar sizes. In Experiment 2, we were able to consider a range of possible sizes because we used three distinct data sets. However, because the size cue is usually out of the visualization designer's control, we chose to focus our evaluation on the effects of spatial proximity and color cues.


\subsection{Participants and Method}
Aside from the recoloring of the bar charts, Experiment 2 followed the same procedures as Experiment 1. We collected sentence conclusions from 114 unique participants ($M_{age} = 39.71$, $SD_{age} = 9.43$, 45 females). Every participant wrote three conclusions (in the order of which came to mind first). After excluding participants who failed the attention check or provided nonsensical conclusions (7.8\%), we ended up with 321 conclusions in total. 

Following the same approach as that in Experiment 1, two coders coded each conclusion into one of the eleven comparison types listed in Table \ref{comparisonTypes}. Overall, the two coders agreed 75.7\% of the time in their ratings, with a high inter-rater reliability Kappa value of 0.716 ($z = 34.6$, $p < 0.001$).

\subsection{Results Overview}

As shown in Figure \ref{exp12ComparisonOverview}, for the first conclusion, participants made group comparisons most frequently (53.6\%), followed by pair comparisons (20.5\%), and a noticeable number of superlative comparisons (11.6\%), all of which occurred significantly more often than the other comparison types (${\chi}^2$ = 124.36, $p < 0.001$). For the second conclusion, participants made pair comparisons most often (36.0\%), followed by group comparisons (30.6\%), and both were significantly more frequent than the other comparisons types (${\chi}^2$ = 59.11, $p < 0.001$). For the third conclusion, participants made pair comparisons most often (36.4\%), followed by group comparisons (25.5\%), and second-order quantity comparisons (22.4\%), all of which occurred significantly more frequently than the other comparison types (${\chi}^2$ = 54.69, $p < 0.001$).

We replicated the finding from Experiment 1 that participants were more likely to make group comparisons as their first comparison and then shifted to make pair comparisons as their second comparison. We also replicated the finding that most of the pair comparisons done were near (48.5\%) and far (44.4\%) pair comparisons and very few were center (4.04\%) and edge (3.03\%) comparisons (${\chi}^2$ = 12.31, $p = 0.031$). We also observed a consistently low frequency of superlative comparisons (11.6\% as first conclusions, 8.1\% as second conclusions, 7.1\% as third conclusions). 

Interestingly, we noticed that the alternating bar colors were able to elicit more second-order comparisons in Experiment 2. Furthermore, although we found no difference between second-order quantity and relation comparisons in Experiment 1, in Experiment 2 their likelihood varied depending on the order in which they were made. A Chi-square analysis with Bonferroni correction for multiple comparisons suggested that viewers were equally likely to make both second-order comparisons as their first and second comparisons (p1 = 0.15, p2 > 0.5), but were more likely to make a second-order quantity comparison as their third comparison (p = 0.08).

\subsection{Hypothesis 5: Grouping Cues in Group Comparisons}
\label{h6}

To compare the strength of spatial proximity and color as grouping cues, we counted the number of times participants conducted a near or a far group comparison. We found that 36.98\% of all group comparisons were far comparisons and 63.03\% were near comparisons. Given the visual design of the bar charts, a near-group comparison would indicate participants are grouping by proximity, whereas a far-group comparison would indicate they are grouping by color. Thus, spatial proximity was a stronger cue than color grouping overall, supporting our hypothesis.

We conducted a post-hoc analysis examining the frequency of participants making near versus far comparisons across the six charts we presented, to see if spatial proximity remained the stronger grouping cue across different data sets (see Figure \ref{exp12groupComparisons}). We used a Chi-square analysis comparing the frequency of participants grouping data values by spatial proximity and color across all six charts. We found that spatial proximity is the dominant grouping cue over color for all but two charts (${\chi}^2$ = 38.15, $p < 0.001$). The first chart is the one that depicts data set 1, grouped by company. Here, participants were \textit{equally likely} to group the values by spatial proximity and color (Bonferroni corrected $p-value = 0.30$. Note that although in Figure \ref{exp12groupComparisons} it may look like near comparisons are made more frequently than far comparisons, the difference is not statistically significant). The second chart depicts data set 3, grouped by company. Here, participants were more likely to group the values by color rather than spatial proximity (Bonferroni corrected $p-value < 0.001$). 

There are two possible explanations for why spatial proximity became the weaker cue for these two charts. The first might be the chart labels. It is possible that people have a linguistic preference to compare A to B more so than East to West. The second possible explanation is that the underlying data values (the size of the bars) might interact with spatial proximity and color, such that size similarities might strengthen the color cues. 

To investigate where there might be a labeling effect, we conducted a follow-up experiment examining whether the chart labels might be the driver of this effect. We recruited an additional 37 participants and followed the same procedure as before. We showed them the chart depicting data set 3, grouped by company, with alternating colors, as shown in the Experiment 2 section of Figure \ref{exp12groupComparisons}, but swapped the axis label of the chart to show revenue of A and B, grouped by region (East and West). We coded participant responses following the exact same procedures. We found that this relabeled chart showed flipped results, such that participants were more likely to make near comparisons than far comparisons (Bonferroni corrected $p-value < 0.001$), indicating that the original results may have been driven by the chart labels. To account for the labeling effect, we averaged the results across the original chart and the relabeled chart and found that 42.5\% of the group comparisons were far comparisons, and 57.5\% of the comparisons were near comparisons, which suggests that spatial proximity still dominates color overall.

In conclusion, spatial proximity is the dominating grouping cue over color cues, but the chart labels might push a viewer toward one grouping or the other depending on the viewer's prior beliefs, expectations, or knowledge. There is no evidence at present that the size of the bars influences grouping, although size was not the focus of the present research.

\subsubsection{Comparing Group Comparisons in Experiment 1 and 2}

We compared the frequencies with which people made near and far group comparisons in Experiment 1 and Experiment 2 to draw more general conclusions about the strength of color grouping cues. In Experiment 1, we found that 40.7\% of the group comparisons were far comparisons and 59.3\% were near comparisons. In Experiment 2, we found that 41.8\% of the group comparisons were far comparisons, and 58.2\% were near comparisons. Chi-square analysis comparing the frequencies of participants making a near comparison versus a far comparison between the two experiments suggests that there is no significant difference in how frequently people made near and far comparisons between the two different coloring schemes (${\chi}^2$ = 0.032, p = 0.86). The recoloring of bars in Experiment 2 did not increase the likelihood of participants making far-group comparisons. Consistent with findings from previous analysis, this suggests that color is a weak grouping cue that has little effect on which bars get grouped together. Figure \ref{exp12groupComparisons}, which compares the frequency of near and far group comparisons in Experiment 1 and Experiment 2, further illustrates that color did not have a strong impact on what groups people formed, as the results from Experiment 1 and 2 are strikingly similar.



\subsection{Hypothesis 6: Grouping Cues in Pair Comparisons}
\label{h7pair}

Just as Experiment 2 allows us to compare spatial proximity and color as grouping cues, it also allows us to compare them as cues for comparison. That is, when participants perform a pair comparison between two bars, are they more likely to compare bars that are beside each other, or bars that are the same color? To examine this question, we looked at the frequency of near-pair comparisons and far-pair comparisons. Overall, in Experiment 2, participants performed near comparisons 48.5\% of the time and far comparisons 44.4\% of the time. 

We took a closer look at how frequently participants made near and far pair comparisons for each of the six charts we showed them in Experiment 2, to compare the strength of spatial proximity and color grouping cues in pair comparisons (see Figure \ref{exp12pairComparison}). We ran a Chi-Square analysis with Bonferroni correction comparing the frequencies with which participants made near and far comparisons for the six charts. The results suggest that overall, participants were \textit{equally likely} to make near and far comparisons except for when they saw the bar chart depicting data set 2, grouped by region (${\chi}^2$ = 12.31, p = 0.019). Again, we tested whether the anomalous finding for data set 2, grouped by region, might be driven by the axis labels by running a follow-up experiment with an additional 39 participants. They viewed the same bar chart, except that the chart had its label switched so it was shown as grouped by company. We found that the effect replicated and participants were still more likely to make far-pair comparisons than near-pair comparisons for this specific dataset and grouping arrangement (${\chi}^2$ = 37.40, $p < 0.001$). 

In conclusion, spatial and color cues are equally salient for pair comparisons within most charts. But why might participants perform differently with data set 2, grouped by region, preferring to compare bars that are the same color over comparing bars that are spatially proximate? We speculate that this effect is driven by the relative \textit{sizes} of the bars. In this chart, the same-color pairs are highly similar in size. Thus, it is possible that participants will tend to select and compare bars that are the same size. 

\subsubsection{Comparing Pair Comparisons in Experiment 1 and 2}

We compared the by-chart frequencies of pair-wise comparisons between Experiment 1 and Experiment 2 to draw a more general conclusion about color as a grouping cue. As shown in Figure \ref{exp12pairComparison}, similar to the group comparisons, the frequencies with which participants made pair comparisons are strikingly similar between the two experiments. The recoloring of the bars in Experiment 2 did not change participants' comparison behavior (${\chi}^2$ = 1.01, p = 0.32). This finding again suggests that color cues in visualization do not significantly influence what people compare, consistent with our findings from Section \ref{h6}. Additionally, considering that color cues have little effect on what people compare, and yet people were equally likely to make near and far pair comparisons, this suggests that spatial proximity also has little effect on which pair comparisons people make. This might imply that bar size plays a bigger role in pair comparisons, consistent with what we found in our investigation of labeling effect in Section \ref{h7pair}.



\begin{figure*}[!ht]
  \includegraphics[width=\linewidth]{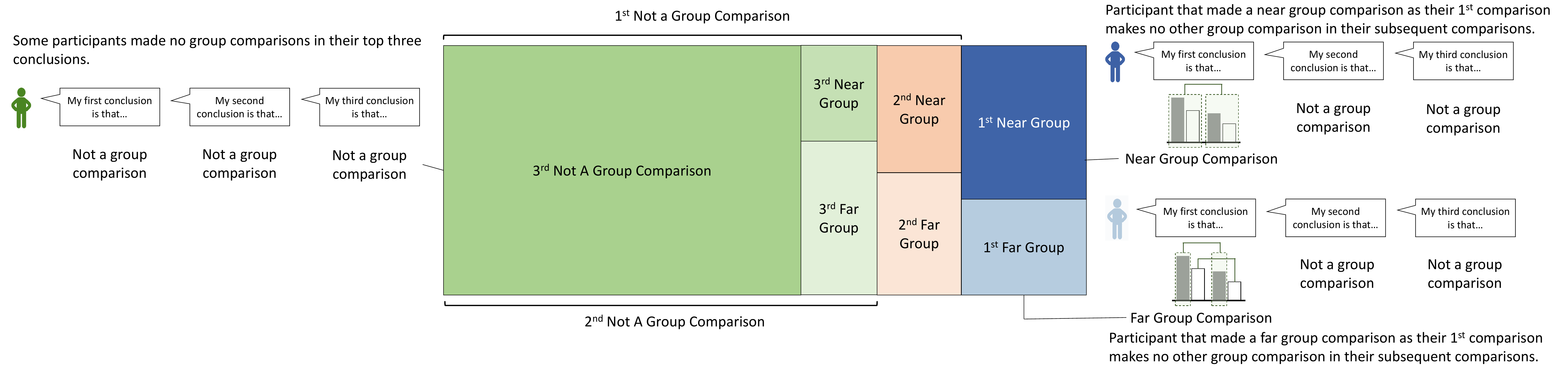}
  \caption{Tree map showing the percentage of people that made near/far group comparisons or no group comparisons for their first, second, and third conclusions. Once the viewer grouped together and compared a set of bars, they no longer consider other possible groupings for the bars.}
  \label{groupComparisonOrder}
  \end{figure*}

\subsection{Hypothesis 7: Shifting between Group Comparisons}
To evaluate Hypothesis 8, which concerns the difficulty of shifting from one grouping to another, we looked at whether participants became less likely to make one group comparison as their second or third conclusion if they had already made the other group comparison as one of their earlier conclusions. 

Across both Experiments, among all the conclusions we collected from participants, 44.7\% of them are group comparisons, with 19.9\% of them being a first conclusion, 12.9\% of them being a second conclusion, and 11.9\% of them being a third conclusion. Although some participants made near comparisons and others made far comparisons, as shown in Figure \ref{groupComparisonOrder}, once a participant made a group comparison of one type, they never made a second group comparison as a later response. 
That is, a participant who made a near-group comparison never made a far-group comparison, and a participant who made a far-group comparison never made a near-group comparison. This finding supports our Hypothesis 8, suggesting that once the viewer groups bars together, they do not consider other possible groupings for the bars. 
This suggests that strong groups could ``constrain'' and pre-attentively guide people's grouping strategies and where people look at, supplementing existing work from psychology on how grouping cues can guide numerical cognition \cite{allik1991occupancy, ciccione2020grouping, halberda2006multiple, franconeri2009number}.

\section{Study Summary}
People may not make the most valuable or informative comparisons when they view a visualization. Instead, they may group, select, and compare elements based on low-level visual features like spatial proximity, color, and size. In the present research, we used established psychological paradigms to study how these visual features influence comparisons when viewing typical bar charts.


We found that, overall, viewers were most likely to make group comparisons early on, followed by pair comparisons. Surprisingly, they rarely made superlative comparisons. They also rarely generated second-order comparisons, corroborating findings in psychology suggesting that second-order comparisons are cognitively difficult to discern and express \cite{halford2005many, shah2011bar}, although they are valuable and often provide more in-depth insights into data. For example, by doing a second-order quantity comparison in Figure \ref{fig:intro}, the viewer can realize that although in general lower noise level and warmer room temperature leads to higher test scores, when the room is noisy, increasing the temperature can improve test scores. In contrast, when the room is quiet, keeping the room cold can improve test scores. 


Spatial grouping cues seemed to dominate color grouping cues, and varying color mapping in bar visualizations had a negligible impact on what comparisons a viewer would make. To summarize, we reflect on whether each of our hypotheses was supported by our empirical data. 

\begin{enumerate} [noitemsep]
  \item  As expected, viewers were more likely to make group comparisons first and pair comparisons second. For the third conclusion, participants again made group comparisons more frequently. 
  
  \item As expected, among pair comparisons, viewers made mostly far and near comparisons. In contrast, center and edge comparisons were rare. 
  
  \item As expected, participants made few second-order comparisons.
  But against our expectations, among second-order comparisons, there was no difference between the number of quantity and relation comparisons. However, the overall small number of second-order comparisons made it difficult to compare the number of each type.
  \item Against our expectations, there were very few superlative comparisons, and people did not tend to seek out those comparisons first.
 
  \item As expected, spatial proximity was the dominant grouping cue. There is no evidence that color or size influences grouping. 
    
  \item Against our expectations, among pair comparisons, viewers were equally likely to compare spatially proximate bars and spatially separate bars. Overall, both spatial and featural cues had little effect on which pairs of bars were selected for comparison. However, there was some limited evidence that participants preferred to select pairs of bars that had a similar size.
  
  \item As expected, people locked into one percept once they made a group comparison and neglected the other group comparison. Those that made a near-group comparison did not make a far-group comparison, and vice versa.
\end{enumerate}

\section{Limitations and Future Directions}
\label{limitation}
We identify several limitations in this study that open up opportunities for future research. First, we relied on three particular sets of underlying data values to serve as a case study for the entire space of 2x2 bar charts, which subsequently served as a case study for the entire space of bar charts, across a number of points, the shape of the data table, and more global aspects of data arrangement (e.g. interleaved vs. small multiple designs \cite{jardine2019perceptual}). Clearly, more work will be needed to confirm how far these results generalize to the larger space of possibilities. Additionally, because we found limited evidence that bar size influenced how people group and compare data values, future research can uncover more specific and generalizable effects of the data value (bar size) by systematically varying the sizes of the bars. In addition to size, there are other grouping features to test, such as bar width and shape. Finally, it is possible we can manipulate the color cues \cite{szafir2017modeling} to rival or beat spatial proximity as a grouping cue. 

Regarding the types of comparisons people make, future researchers can further explore why few participants made superlative comparisons. As mentioned before, findings from perceptual psychology literature suggest that superlatives are easy to identify. Participants might have spotted the superlatives but chosen not to mention them because they were simply too `obvious' of a conclusion. 

The verbal description method used in the present study is inspired by literature on `sentence-picture congruity' from psychology \cite{clark1972process}, and we used it as a proxy for the visual comparison that was most salient to a viewer. Future researchers can explore other ways to elicit viewer conclusions from visualizations, such as asking what comparisons people remember after a brief viewing, what changes to data values they notice, or what decisions they make based on the data. By gathering converging evidence across these disparate measures, this line of research can help the community more effectively elicit viewer percepts from visualizations. 

We were not able to make detailed observations about how people perform second-order comparisons in the present experiment, due to the fact that very few people made these types of comparisons. This finding suggests that, although scientific and news articles often use these bar charts to communicate findings and expect the audiences to make second-order comparisons from them, bar charts are not effective in eliciting these more complex comparisons. Future researchers could explore other visual representations that may better afford second-order comparisons and examine how visual grouping cues could help.
\vspace{2mm}

\begin{figure}[t]
 \includegraphics[width=\columnwidth]{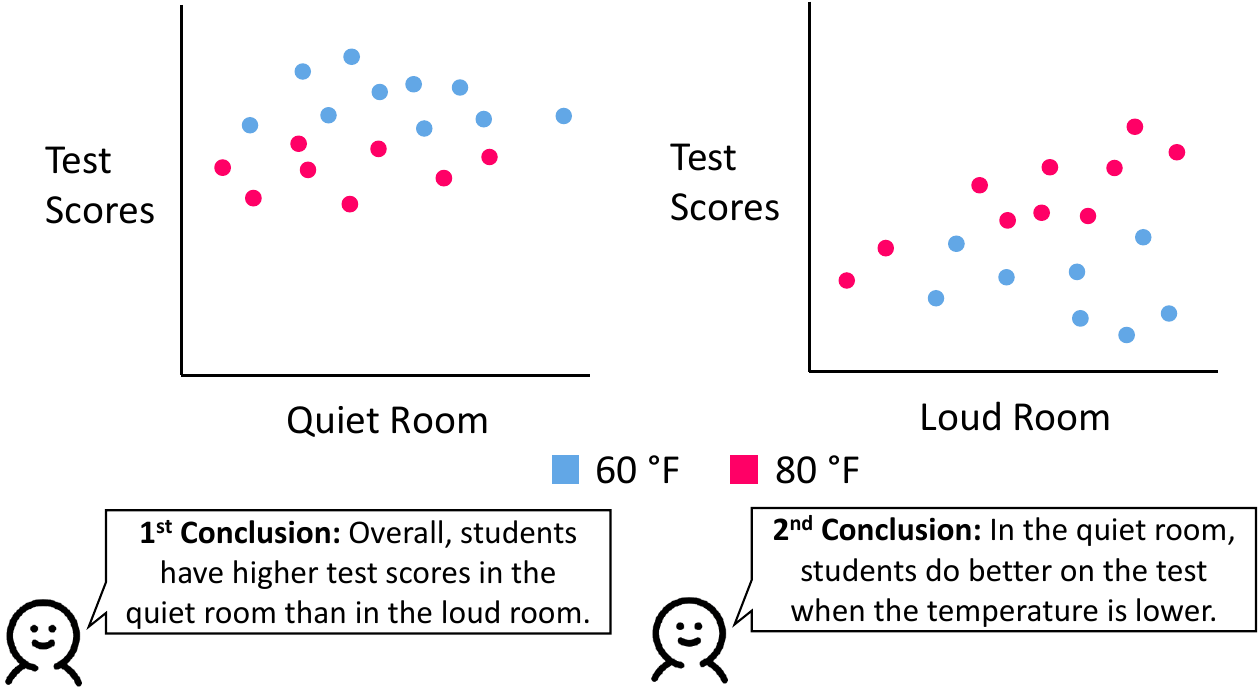}
 \caption{Example showing how the comparison tendencies identified in the present studies can be extended to apply to other types of visualizations. People are likely to prioritize spatial grouping cues, grouping the points by room and comparing the two rooms, and then comparing the two colors within the same room.}
 \label{Generalizability}
\end{figure}

\noindent \textbf{Generalizability:} Although the experiments in the paper tests bar charts as a case study, we believe there are strong reasons to predict the results will generalize, as the psychological and visual concepts applied in this paper originated in other areas of research, including visual intelligence tests (e.g., \cite{lovett2017modeling, jardine2019perceptual}). The experiments we conducted so far corroborate with similar experiments using very different stimuli (e.g., abstract shapes \cite{yu2019similarity}), which demonstrates that these concepts can generalize across two very different visual domains. Figure \ref{Generalizability} illustrates an example of how the tendency to make group comparisons based on the spatial grouping cues might apply to a different chart type - scatterplots. Future work can extend this investigation to test viewers' comparison tendencies in more complex chart types with more complex tasks, to potentially identify boundary conditions in which these findings may no longer generalize.
\vspace{2mm}

\noindent \textbf{Top-Down vs. Bottom-Up:} The charts shown in the present work omitted titles, annotations, and any other context that might guide comparison. This is a deliberate design choice to control for the viewers' goals and expertise in order to maximally isolate the impact of the graph design through bottom-up processing guided by spatial and color cues. This makes our investigation more closely model how novices explore visualizations. As novices typically do not know the appropriate strategies nor possess the appropriate information to know where to look, we conjecture that visual attention will be driven primarily by bottom-up factors, such as a tendency to group adjacent objects together and attend to them as a single object. 

Despite this, we found an effect of how the charts are labeled (A/B vs. East/West) on what people ultimately compare. This finding suggests that people's prior knowledge, experiences, and goals are critical to consider in visualization design, and future work could assess their influence against that of the chart's design. For example, future work can explore how people with varying levels of expertise explore visualizations by incorporating top-down processing mechanisms in their investigation. 
Top-down processing mechanisms suggest that spatial attention can be shifted deliberately to a target location in an image when the viewer knows where to look \cite{baluch2011mechanisms}. Similarly, attention can be guided towards a target visual feature, for example, a color or a shape, when the viewer knows what to look for. These effects are important when an expert views a graph, as the expert can strategically focus attention on the parts of the graph that are most relevant for extracting a piece of information. 

Looking ahead, researchers should consider examining the both effects of bottom-up (i.e., stimulus-driven) and top-down (i.e., knowledge or task-driven) effects in combination when conducting future studies at the intersection of human perception/cognition and data visualization. 
\vspace{2mm}

\noindent \textbf{Visual Literacy:} Existing work has demonstrated that individual preferences have an effect on how people interact with visualizations \cite{stokes2022more}.
While the present paper was not designed nor powered to identify the effects of individual differences such as visual literacy and expertise, we acknowledge that by studying both visualization novices and experts, the visualization community can gain a richer understanding of how people strategize when inspecting and interpreting visualizations. 
Future work can deploy larger-scaled studies to further investigate how the amount of experience participants have with visualizations, which can include expertise with reading visualizations more generally, expertise with the domain of data being presented, or the amount of education received regarding charting conventions and statistical inferences, changes people's grouping strategies and chart takeaways. 
\vspace{2mm}

\noindent \textbf{Linguistic Ambiguities:} In our qualitative coding, we encountered cases where the participants provided ambiguous responses that were difficult to categorize confidently into comparisons.
For example, amongst the second-order comparisons made, one participant wrote ``they are opposite'' and another wrote ``revenue is similar.''
The coders had disagreements on which second-order comparison category they fit under (i.e., relation or quantity).
Future work could consider alternative ways to elicit participant responses when describing chart takeaways to further disambiguate participants' intentions, such as asking participants for drawing as described in \cite{xiong2021visual} and \cite{kim2019bayesian}, or through eye-tracking (or equivalent alternatives) studies to leverage gaze shifts to determine which values people attended to as described in \cite{shin2022scanner, kim2015crowdsourced, bylinskii2017eye, kim2017bubbleview}. 

Furthermore, future work could also employ a different prompt to elicit chart takeaways from participants. 
The present experiment used the word ``conclusion'' to elicit a more generic response, but alternative approaches such as explicitly asking for comparisons or key takeaways might lead to the unveiling of perceptual rules that apply when people interact with visualizations. 
Future work can also examine the effect of top-down processing and task demands to ask participants to either complete specific chart comprehension or data trend prediction tasks (e.g., as in \cite{mantri2022viewers} and \cite{burns2020evaluate}.  
\vspace{2mm}

\noindent \textbf{Applications to Natural Language Visualization Tools:} Finally, our data set comprises hundreds of verbal descriptions of bar visualizations. This data set could be expanded in both volume and scope of visualization types, and text descriptions could be augmented with automated mining of existing descriptions. 
If a natural-language processing algorithm (or crowdworkers) could be leveraged to turn these descriptions into the type of comparisons shown in Table \ref{comparisonTypes}, and the visualizations could be discretized into qualitative descriptions, (e.g. Vega \cite{satyanarayan2016vega}), the combined data set could become the input that trains a more comprehensive prediction engine for what comparisons are made by a typical visualization viewer (e.g.,\cite{Srinivasan2019, srinivasan2021collecting}). This engine can also produce text descriptions of a visualization, which helps make visualizations more accessible for visually-impaired users (see \cite{jung2021communicating} for more).

\section{Design Guidelines}
\label{guidelines}
Based on our results, we can propose design guidelines for visualization designers and visualization recommendation systems that will help viewers more efficiently extract insights from data. Designers should place chart elements close together if they want viewers to group those elements and place the elements far apart if they do not want the viewer to group them---visual features like color and size may also influence the likelihood of grouping, but our research suggests proximity is by far the dominant cue, at least within bar charts. We believe it is critical that designers be aware of these grouping cues because they determine what comparisons are available to a viewer. After a viewer groups elements together, there will be a delay before the viewer is able to compare the individual elements. On the other hand, the viewer will likely never even consider a comparison that relies on a different grouping from the one they originally formed, due to the difficulty of perceptual reorganization.

Most current work on visual comparison focuses on the efficiency or precision of a given comparison. We argue that these factors are important, but irrelevant if a visualization viewer never makes the appropriate comparison in the first place. In our real-world experience, the biggest delays in understanding a new visualization are knowing how to read a visualization and knowing which patterns one should pay attention to. When communicating data, analysts tend to assume that the viewer sees what they see \cite{xiong2019curse}, instead of designing the visualization to push the viewer toward seeing the `right' pattern. We hope that refinement of the type of model developed here can lead to a set of guidelines or concrete tools to help analysts achieve that goal. 


\bibliographystyle{abbrv}
\bibliography{comparison}

\pagebreak 

\begin{wrapfigure}[43]{l}[0mm]{25mm} 
    \includegraphics[width=1in,height=1in,clip,keepaspectratio]{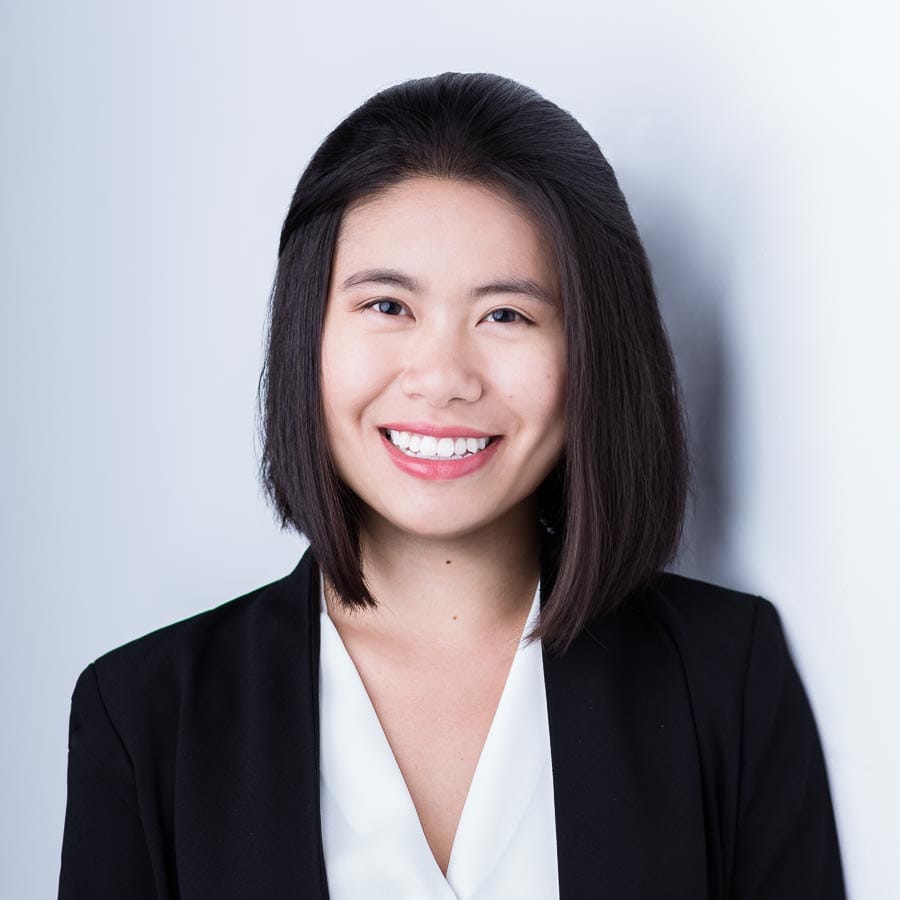}\par
    \vspace{5pt}
    \includegraphics[width=1in,height=1in,clip,keepaspectratio]{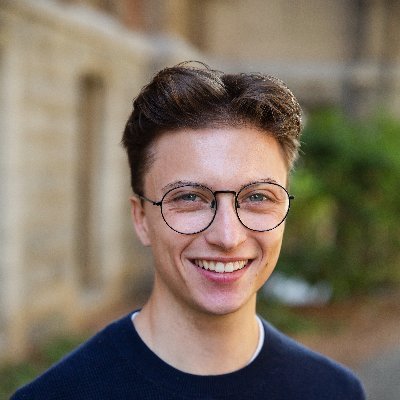}\par  
    \vspace{5pt}
    \includegraphics[width=1in,height=1in,clip,keepaspectratio]{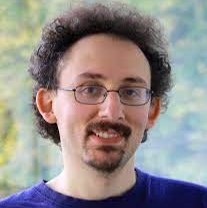}\par
    \vspace{5pt}
    \includegraphics[width=1in,height=1in,clip,keepaspectratio]{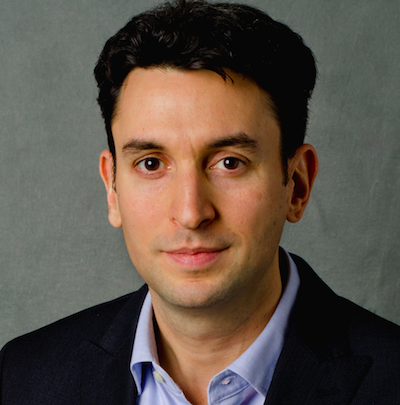} 
  \end{wrapfigure}
  \textbf{Cindy Xiong Bearfield} is an incoming Assistant Professor in the School of Interactive Computing at Georgia Tech. By investigating how humans perceive and interpret visualized data, she aims to improve visualization design, data storytelling, and data-driven decision-making. \par
  \vspace{9pt}
  \textbf{Chase Stokes} is a graduate student at UC Berkeley School of Information. He studies information visualizations and text, with particular interest in improving design practices for combining visual and textual information.\par
  \vspace{28pt}
  \textbf{Andrew Lovett} is a cognitive scientist at the US Naval Research Laboratory. He studies visual attention and perception in humans and machines. \par
  \vspace{38pt}
  \textbf{Steven Franconeri} is a Professor of Psychology at Northwestern University, and Director of the Northwestern Cognitive Science Program. He studies visual thinking and visual communication, across psychology, education, and information visualization.\par

\end{document}